\documentclass[prb,twocolumn,showpacs,amsmath,amssymb,nobibnotes,citeautoscript,floatfix,superscriptaddress]{revtex4}

\usepackage{graphicx}
\usepackage{amsmath}
\usepackage{amssymb}
\usepackage{bm}

\begin{document}

\title{A quasi-local measure of inter-scale transfer: An approach to
  understanding turbulence}

\author{M.K.~Rivera} \affiliation{Condensed Matter \& Thermal Physics
  Group,Los Alamos National Laboratory, Los Alamos, NM 87545}
\affiliation{Center for Nonlinear Studies, Los Alamos National Laboratory, Los
  Alamos, NM 87545}

\author{W.B.~Daniel} \affiliation{Condensed Matter \& Thermal Physics
  Group,Los Alamos National Laboratory, Los Alamos, NM 87545}
\affiliation{Center for Nonlinear Studies, Los Alamos National Laboratory, Los
  Alamos, NM 87545}

\author{S.Y.~Chen} \affiliation{Center for Nonlinear Studies, Los Alamos
  National Laboratory, Los Alamos, NM 87545} \affiliation{Department of
  Mechanical Engineering, The Johns Hopkins University, Baltimore, MD 21218}

\author{R.E.~Ecke} \affiliation{Condensed Matter \& Thermal Physics Group,Los
  Alamos National Laboratory, Los Alamos, NM 87545} \affiliation{Center for
  Nonlinear Studies, Los Alamos National Laboratory, Los Alamos, NM 87545}

\date{\today}

\begin{abstract}
  Many questions remain in turbulence research---and related fields---about
  the underlying physical processes that transfer scalar quantities, such as
  the kinetic energy, between different length scales. Measurement of an
  ensemble-averaged flux between scales has long been possible using a variety
  of techniques, but instantaneous, spatially-local realizations of the
  transfer have not. The ability to visualize scale-to-scale transfer as a
  field quantity is crucial for developing a clear picture of the physics
  underlying the transfer processes and the role played by flow structure. A
  general technique for obtaining these scale-to-scale transfer fields, called
  the filter approach, is described.  The effects of different filters, finite
  system size, and limited resolution are explored for experimental and
  numerical data of two-dimensional turbulence.
\end{abstract}

\maketitle
\section{Introduction \label{sec: Introduction}}

Turbulence is characterized by the transfer of an inviscid constant, such as
kinetic energy or a passive scalar, between different length scales. One of
the primary goals of turbulence research is to understand the mechanisms that
drive this transfer process. Numerical simulations performed by Farge {\it et
  al.}~suggest that in both two-dimensional (2D) and three-dimensional (3D)
turbulence it is a relatively small number of ``coherent structures'' that
dominate the turbulent dynamics \cite{Farge_1999_PhF,Farge_2001_PRL}. As a
result, it is important to understand how the existence and interaction of
these coherent structures affect the transfer processes. Thus, the {\it
  spatially-local} transfer properties of the flow need to be measured and
correlated with these structures.

In this manuscript a tool for obtaining local information about the
scale-to-scale transfer of inviscid constants from experimental/numerical data
is examined.  The method, called the ``filter approach'' (FA), has
traditionally been applied to large-eddy numerical simulations (LES)
\cite{Lesieur_Turbulence}, but is developed here in the context of
experimental data analysis. By applying a low-pass spatial filter to the
equations of motion (the incompressible Navier-Stokes equation), separate
equations for the filtered, or large-scale, fields and the remaining
small-scale fields can be derived \cite{Germano_1992_JFM}. Within the
resulting equations are coupling terms that represent the interaction of the
large- and small-scale fields with each other. In LES schemes the coupling
terms in the large-scale equations are modeled thereby eliminating the
necessity of directly computing the small scales \cite{Lesieur_Turbulence}. In
our application of the filter approach, however, data from direct numerical
simulations (DNS) or high resolution experiments are used to directly evaluate
the coupling terms and obtain a quasi-local measure of the inter-scale
transfer \cite{Tao_2002_JFM,Rivera_2003_PRL}.

For readers familiar with LES and turbulence modeling it should be stressed
that the philosophy driving our use of the filter approach is different
from the {\it a priori} development of LES models discussed in
\cite{Menevau_2000_ARFM}.  There the objective was to determine empirically
which of several LES modeling schemes for the large- to small-scale coupling
terms most successfully emulates physical data.  In that case, FA was used
primarily as a benchmark, and only went as far as measuring the inter-scale
coupling term.  Rather than producing results for numerical benchmarking, FA
can also be used as an analysis probe to determine where and when in a flow
scale-to-scale transfer of inviscid constants takes place.  In this way one
can isolate important interaction events and form an understanding of
turbulence inter-scale transfer mechanisms.  Of course, this understanding
could eventually be incorporated into LES models.

The filter approach is applied in this paper to both experimental and
numerical data. The intent is not to investigate the underlying physics of the
turbulence, which will be presented in later papers, but rather to determine
the appropriate interpretation of the results and the limitations imposed by
different filters, spatial boundaries, and finite measurement resolution.

\section{Experimental and Numerical Systems \label{sec: Experimental}}

\begin{figure}
  \includegraphics[width=3.2in]{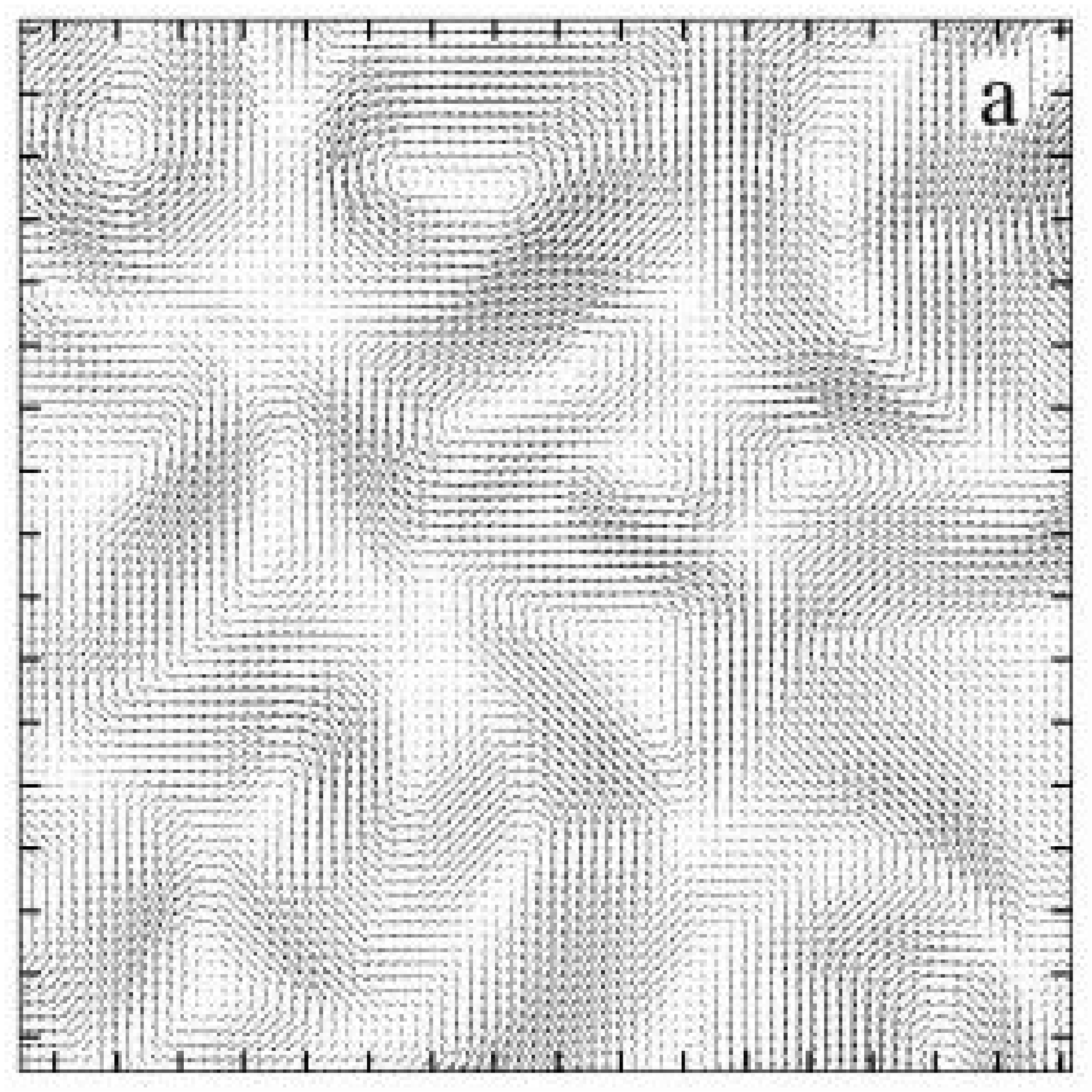}
  \vskip 0.05in
  \includegraphics[width=3.2in]{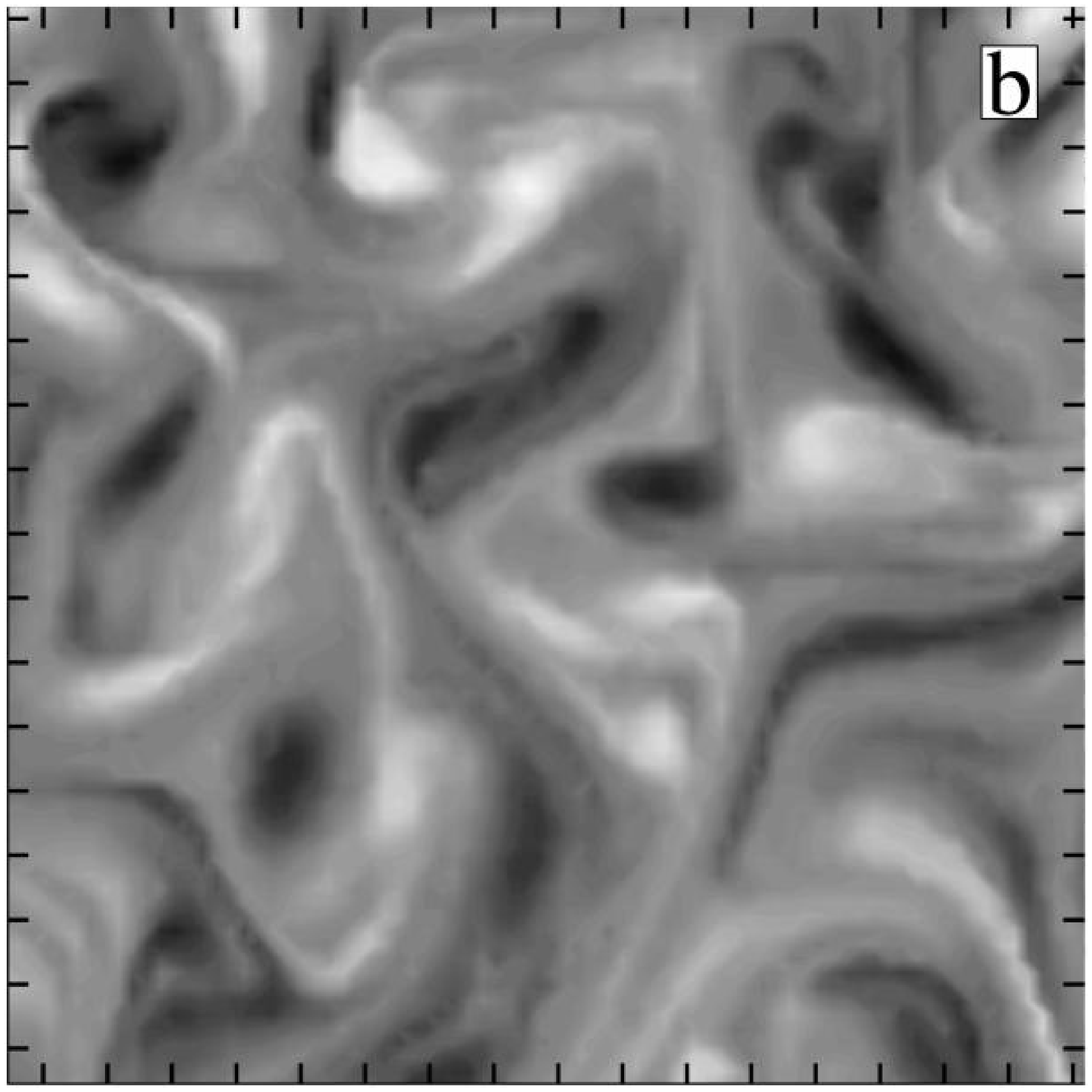}
   \caption{Typical a) velocity and corresponding b) vorticity field
     obtained from the flowing soap film channel.  The hash marks represent
     $1$ mm increments, the mean flow (subtracted out) is in the $-\hat{y}$
     direction.  The top of the image is $\approx 3$ cm downstream from the
     energy injection grid.}
\label{fig: typical-fields}
\end{figure}

Experimental measurements were carried out in a flowing soap-film channel, a
quasi-2D system in which decaying turbulence of low to moderate Reynolds
number can be generated ($10^2 \leq Re \leq 10^4$).  The channel was $5$ cm
wide and was inclined at an angle of $75^{\circ}$ with respect to vertical.
The mean flow was $120$ cm$/$s and the film thickness was about $10$ $\mu$m.
A more detailed description of the channel can be found in
\cite{Rivera_1998_PRL,Vorobieff_1999_PhF}.  Using the empirical relationships
measured in \cite{Vorobieff_1999_PRE}, the films' kinematic viscosity was $\nu
\approx 0.03$ cm$^2$/s.  The turbulence generating grid consisted of rods of
$0.12$ cm diameter with $0.22$ cm spacing between the rods.  Thus, the
blocking fraction is around $0.3$, which is typical for turbulence in 2D soap
film flows
\cite{Gharib_1989_PhD,Kellay_1995_PRL,Rivera_1998_PRL,Rutgers_2001_RSI}.  The
resulting Reynolds number, $Re = UL/\nu$, was $880$ based on the mean-flow
velocity and an injection scale of $L_{\text{inj}} = 0.22$ cm.  The turbulent
velocity, ${\mathbf u}({\mathbf x})$, and vorticity, $\omega({\mathbf x})$,
fields generated by the grid were obtained by tracking $3-5$ $\mu$m
polystyrene spheres (density approximately $1.05$ g/cc) within a $1.8 \times
1.8$ cm$^2$ region located $6$ cm downstream from the grid (20-30 eddy
rotation times)\cite{Ishikawa_2000_MST,Ohmi_2000_EiF}.  The particles were
illuminated with a double pulsed Nd:Yag laser and their images captured by a
12-bit, $2048\times2048$ pixel camera.  Around $3\times10^4$ particles were
individually tracked for each image pair and their velocities and local shears
were interpolated to a discrete $135 \times 135$ grid. One-thousand velocity
and vorticity fields were obtained in this way and were used to compute
ensemble averages of the statistical measures described below. Typical
velocity and vorticity fields are shown in Fig.~\ref{fig: typical-fields}.

To supplement the experimental data, a direct numerical simulation of the 2D
Navier Stokes equation was performed. Computational details are presented in
\cite{Chen_2003_PRL}.  The equation
\begin{equation}
\frac{\partial \omega}{\partial t} + u_j \frac{\partial \omega}{\partial x_j}
+\nu_m (-1)^m \frac{\partial^{2m} \omega}{\partial x_j^{2m}} = F
\end{equation}
was simulated in a square domain with side $L=2\pi$ and periodic boundary
conditions. Here, $u_i$ is the $i$-th component of the velocity;
$\omega=\epsilon_{ij}\partial_j u_i$ is the vorticity; and $F$ is a stirring
force applied to wave numbers $|{\bf k}|=4 \rightarrow 7$. The Einstein
summation convention is used throughout. Two values of $m$ are
considered: $m = 1$ corresponding to Laplacian viscosity ($\nu_1 = 0.01$),
and $m = 8$ corresponding to hyper-viscosity ($\nu_8 = 1.4 \times
10^{-7}$), which has the effect of extending the inertial range.

\begin{figure}[b]
   \includegraphics[width=3.2in]{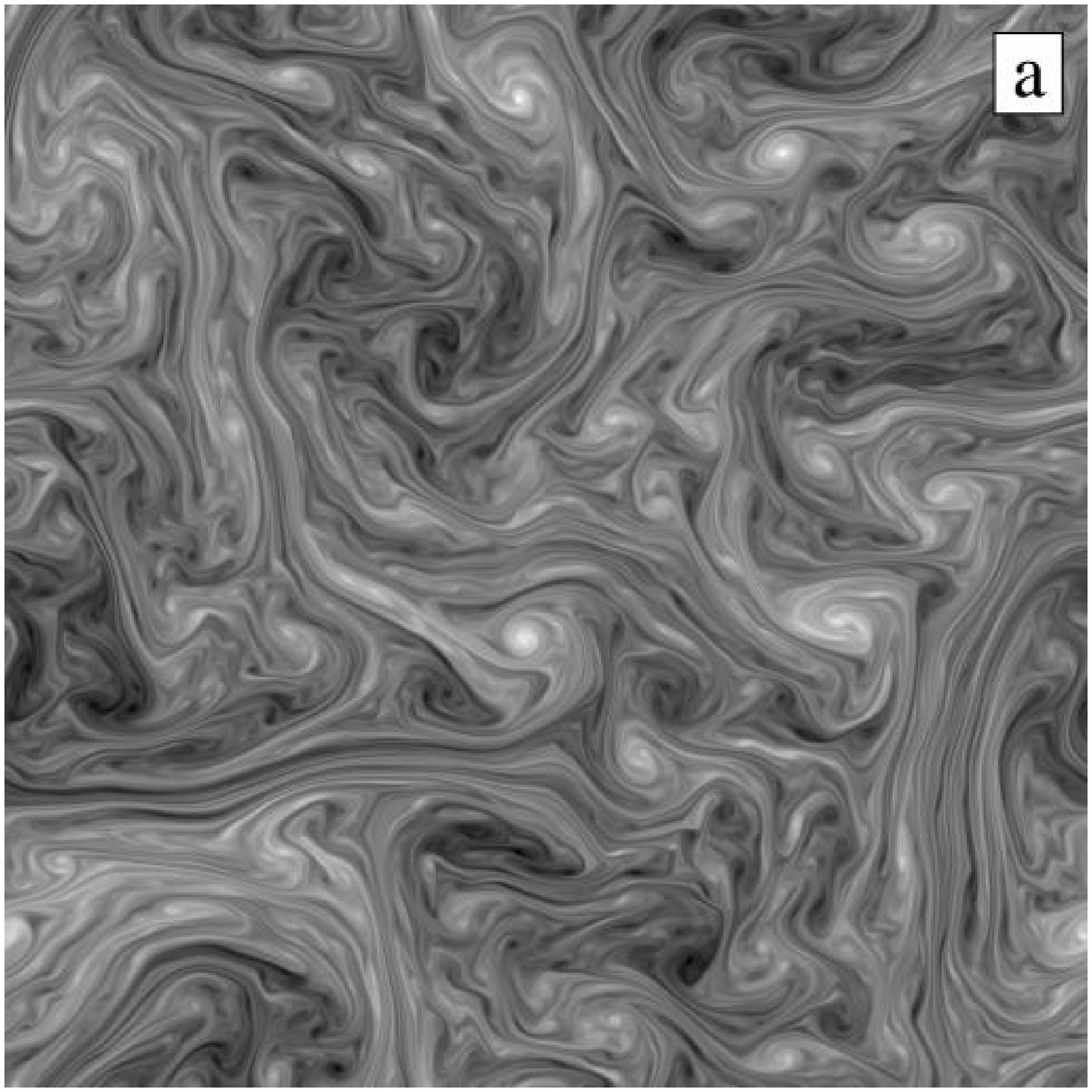} \vskip 0.05in
   \includegraphics[width=3.2in]{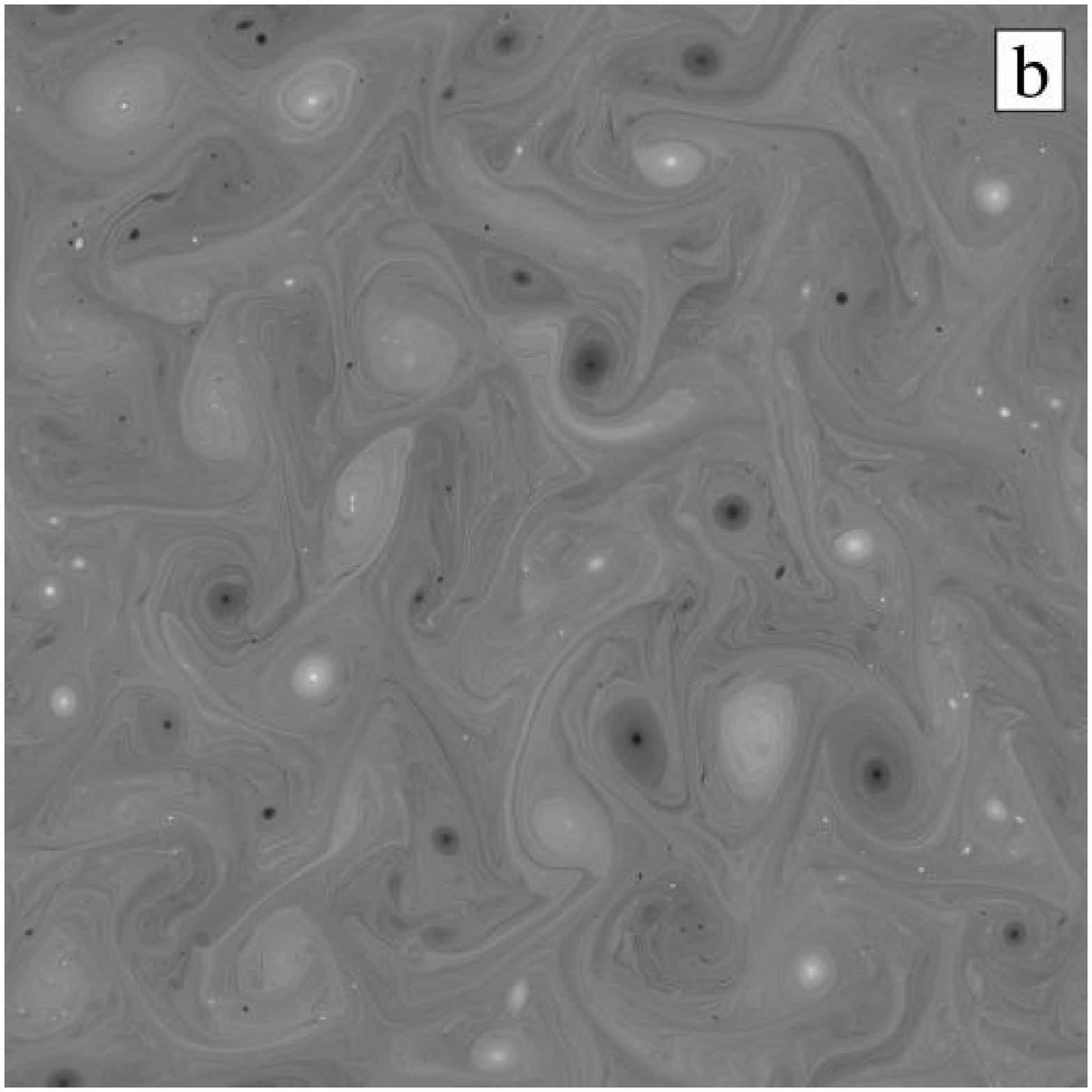}
   \caption{DNS vorticity fields with a) Laplacian viscosity ($m = 1$) and b)
   hyper-viscosity ($m = 8$)}
\label{fig: numerical-fields}
\end{figure}

The equation was solved using a fully de-aliased, parallel pseudo-spectral code
with second-order Adam-Bashforth time-stepping. The resolution was $2048^2$. A
statistically-stationary state was achieved after about 200 large-eddy
turn-over times.  Representative examples of vorticity fields generated using
Laplacian and hyper-viscosity are shown in Fig.~\ref{fig: numerical-fields}.

\section{The Filter Approach \label{sec: FA}}

\subsection{General Application \label{subsec: General Application}}

The filter approach allows the direct measurement of the coupling between
scales in non-linear systems.  It was originally developed as a tool to
truncate numerical simulations of turbulence by modeling small-scale behavior
with the knowledge of large-scale behavior, {\it i.e.} LES
\cite{Lesieur_Turbulence}. In this paper, the general application of FA to
nonlinear systems will be considered first, followed by a specific example:
the case of energy and enstrophy transfer between length scales in 2D
turbulence.

General features of the transfer process are examined by considering a field,
$q({\mathbf x})$, that is evolved by a non-linear evolution operator,
$F^{(\text{nl})}$, as $F^{(\text{nl})} q =0$.  Given a scale, $l$, one can
separate the field, $q$, into large-scale, $q_l$, and small-scale, $q_s$,
components.  This is done by convolving a filter function, $G_{l}$, with $q$
and defining $q_l \equiv G_{l} \ast q$ and $q_s \equiv q - q_l = (\delta -
G_{l}) \ast q$, where $\delta$ is the Dirac delta function.  Applying $G_l$ to
the evolution equation for $q$ yields $G_{l} \ast (F^{(\text{nl})} q)=0$.
Adding zero, written as $F^{(\text{nl})}q_l - F^{(\text{nl})}q_l$, results in:
\begin{equation}
F^{(\text{nl})}q_l = -(G_{l} \ast (F^{(\text{nl})} q) -
F^{(\text{nl})}q_l).
\label{eq: general-filter-space}
\end{equation}
Eq.~(\ref{eq: general-filter-space}) indicates that the large-scale field,
$q_l$, evolves in exactly the same manner as the full field, $q$, up to a
coupling term $C(q_l,q)\equiv-(G_{l} \ast (F^{(\text{nl})} q) -
F^{(\text{nl})}q_l)$.  This coupling term, which can be considered as an
external forcing or damping, arises from the small-scale field, $q_s$,
interacting with $q_l$. Thus, $C$ provides a measure of the interaction
between large and small scales. The measurement of $C$ requires no assumptions
about the field, $q$, such as homogeneity or isotropy.  Moreover, $C$ is a
{\em field} quantity, not simply an average, and can reveal not only
information about the coupling between scales, but also when and where such
interactions are taking place and with what strength.

The field nature of the coupling term derived above, $C$, makes FA very
valuable in the context of studying turbulence. As mentioned in the
introduction, turbulence is characterized by an average scale-to-scale flux of
quantities such as the kinetic energy. Many possible mechanisms underlying the
transfer processes have been suggested, {\it e.g.}, the stretching of vortex
tubes into thin filaments may account for some fraction of the down-scale
transfer of energy in 3D \cite{Frisch_Turbulence}. In 2D, vortex merger has
been postulated as a way of transferring energy to larger scales
\cite{Lesieur_Turbulence}. The difficulty is that neither of these pictures
has been conclusively correlated with topological flow structures or the
underlying transfer dynamics, though some attempts have been made
\cite{Daniel_2002_PRL}.  By using FA to measure the inter-scale transfer and
correlating this field with the position of flow structures (identified by
other means), one can determine the veracity of these hypothesized transfer
mechanisms.

There are a number of subtleties to consider when applying FA to experimental
data.  First, there is the choice of the filter function, $G_l$.  The
interpretation of the scale-to-scale transfer depends on the selection of the
filter function.  For example, if the convolution function is Gaussian defined
by length scale $l$, the interpretation of the filtered functions is as given
above.  On the other hand, if the convolution is with the kernel $H_l \equiv
(\delta - G_l)$, where again $G_l$ is Gaussian and $\delta$ is a Dirac delta
function, the resulting interpretation of the convolved fields, $H_l \ast f$,
as ``large-scale'' is incorrect.  Indeed, the function $H_l$ produces
``small-scale'' fields.  Other ramifications of changing the filter function
will be explored shortly but the convention is adopted that the filter
function is always low-pass.

Two additional considerations when applying FA to real data are finite
measurement resolution and the existence of spatial boundaries.  The extent to
which the measurements are sensitive to either of these factors depends on the
quality of the data and on the form of the filter used.  Significant research
in the context of LES has addressed similar concerns, but always for the
purpose of approximating the physical system in a numerical simulation
\cite{Lesieur_Turbulence}. These issues are explored in some depth in a later
section.

\subsection{Application to two-dimensional turbulence: energy and enstrophy
  transport \label{subsec: 2D Application}}

In two-dimensional turbulence kinetic energy, $E \equiv u^2/2$, and enstrophy,
$\Omega \equiv \omega^2/2$, are conserved in the inviscid limit. Theory,
numerics, and experiments all indicate that energy is transferred on {\it
  average} from small to large length scales (up-scale) and that enstrophy is
transferred in the opposite direction (down-scale) \cite{Kraichnan_1980_RPP}.
FA, however, can be used to obtain much more detailed information about the
energy and enstrophy transfer within the flow.  We begin by examining the
scale-to-scale coupling of energy in the 2D Euler equation (viscous terms are
linear and, hence, cause no direct scale-to-scale transfer, eliminating the
need to examine the full Navier-Stokes equation):
\begin{equation}
\frac{\partial u_i}{\partial t} + u_j \frac{\partial u_i}{\partial
x_j} = -\frac{\partial p}{\partial x_i},
\label{eq: Euler-Velocity}
\end{equation}
where $u_i$ is the $i$-th component of the velocity field, $p$ is the density
normalized pressure field, and summation over repeated indices is assumed.
Contracting this evolution equation with a filter function, $G_l$, and
extracting the coupling term as described above yields:
\begin{eqnarray}
\frac{\partial (u_i)_l}{\partial t}&+&(u_j)_l \frac{\partial 
(u_i)_l}{\partial
   x_j} \nonumber \\&=& -\frac{\partial p_l}{\partial x_i} - 
\frac{\partial}{\partial x_j}((u_i u_j)_l
- (u_i)_l(u_j)_l).
\label{eq: Large-Scale-Euler-Velocity}
\end{eqnarray}
The notation $(f)_l$, or simply $f_l$, will be used to denote the large-scale
field $G_l \ast f$.  Eq.~(\ref{eq: Large-Scale-Euler-Velocity}) is almost the
equivalent of Eq.~(\ref{eq: general-filter-space}).  The one delicacy is that
the term $p_l$ is not the large-scale pressure field, {\it i.e.}, it is not
the field obtained using the gradients of $(u_i)_l$.  This is not an important
issue here, however, because, as will be demonstrated later, $p_l$ does not
contribute to the inter-scale transfer.

From Eq.~(\ref{eq: Large-Scale-Euler-Velocity}) the scale-to-scale coupling
term for the velocity is given by
\begin{equation}
C((u_i)_l,u_i) = -\frac{\partial \tau^{(l)}_{ij}}{\partial x_j},
\end{equation}
where $\tau^{(l)}_{ij} = (u_i u_j)_l - (u_i)_l (u_j)_l$ is the subgrid-scale
stress tensor. To obtain the equation for the evolution of large-scale energy,
$E^{(l)}=(u_i)_l^2/2$ one multiplies Eq.~(\ref{eq:
  Large-Scale-Euler-Velocity}) by $(u_i)_l$.  (Note that the notation
$E^{(l)}$ is used rather than $E_l$ since $E_l = (u_i^2)_l/2 \neq (u_i)_l^2/2
= E^{(l)}$.)  The resulting equation,
\begin{eqnarray}
\frac{\partial E^{(l)}}{\partial t} + \frac{\partial}{\partial 
x_j}((u_j)_l
E^{(l)} + (u_j)_l p_l) = -(u_i)_l \frac{\partial 
\tau^{(l)}_{ij}}{\partial x_j}.
\label{eq: Large-Scale-Energy-1}
\end{eqnarray}
for the energy contained at scales larger than $l$, is identical to the full
energy evolution equation up to the coupling term on the right hand side
(again ignoring the pressure term).

This coupling term is not yet in the form to directly yield scale-to-scale
energy transfer information.  There are two ways in which the small-scale
velocities can change the large-scale energy: by physically transporting it
from point-to-point or by transferring it between scales. To separate the two,
the Leibnitz rule is used to rewrite the right hand side of Eq.~(\ref{eq:
  Large-Scale-Energy-1}) as
\begin{equation}
=-\frac{\partial}{\partial x_j} ((u_i)_l \tau^{(l)}_{ij}) +
\tau^{(l)}_{ij}\frac{\partial (u_i)_l}{\partial x_j}.
\end{equation}
Notice that the latter of these two terms is Galilean invariant, whereas the
former is not.  Boosts to the reference frame should not change the
scale-to-scale transfer of energy but will change the point-to-point
transport.  Therefore, the former term is attributed to the point-to-point
transport and the latter to the scale-to-scale transfer.

Another way to contrast the point-to-point coupling with scale-to-scale
coupling is to consider the limit of a homogenous system.  In this case taking
an ensemble average, $\langle \ldots \rangle$, should eliminate all
point-to-point transport terms, leaving only inter-scale transfer
contributions.  Since the action of ensemble averaging commutes with the
derivative operation, and since the spatial derivative of an ensemble average
is zero in the limit of homogeneity, the ensemble average of Eq.~(\ref{eq:
  Large-Scale-Energy-1}) is simply $ \partial_t \langle E^{(l)} \rangle =
\langle \tau_{ij}^{(l)}\partial_j (u_i)_l \rangle.$ Not only does this
demonstrate that $\partial_j[\tau^{(l)}_{ij}(u_i)_l]$ is a point-to-point
term, but it also demonstrates that all of the terms on the left, other than
the time derivative, are point-to-point as well.  This fact allows us to
ignore the delicacy with respect to the pressure term: it does not affect
scale-to-scale transfer.

For simplicity Eq. \ref{eq: Large-Scale-Energy-1} is rewritten as
\begin{equation}
\frac{\partial E^{(l)}}{\partial t} + \frac{\partial 
J^{(l)}_j}{\partial x_j}
= -\Pi^{(l)},
\label{eq: Large-Scale-Energy}
\end{equation}
where $J^{(l)}_j$ and $\Pi^{(l)}$ are defined as
\begin{eqnarray}
J^{(l)}_j& \equiv& (u_j)_l E^{(l)} + (u_i)_l \tau^{(l)}_{ij} + (u_j)_l p_l
\label{eq: J-def},\\
\Pi^{(l)}& \equiv& -\tau^{(l)}_{ij}\frac{\partial (u_i)_l}{\partial 
x_j} = -\tau^{(l)}_{ij}(S_{ij})_l
\label{eq: Q-def}.
\end{eqnarray}
where $(S_{ij})_l$ is the large-scale strain tensor.  The negative sign in the
definition of $\Pi^{(l)}$ is added so that down-scale transfer has a positive
value whereas up-scale transfer is negative.

\begin{figure*}
   \includegraphics[width=7in]{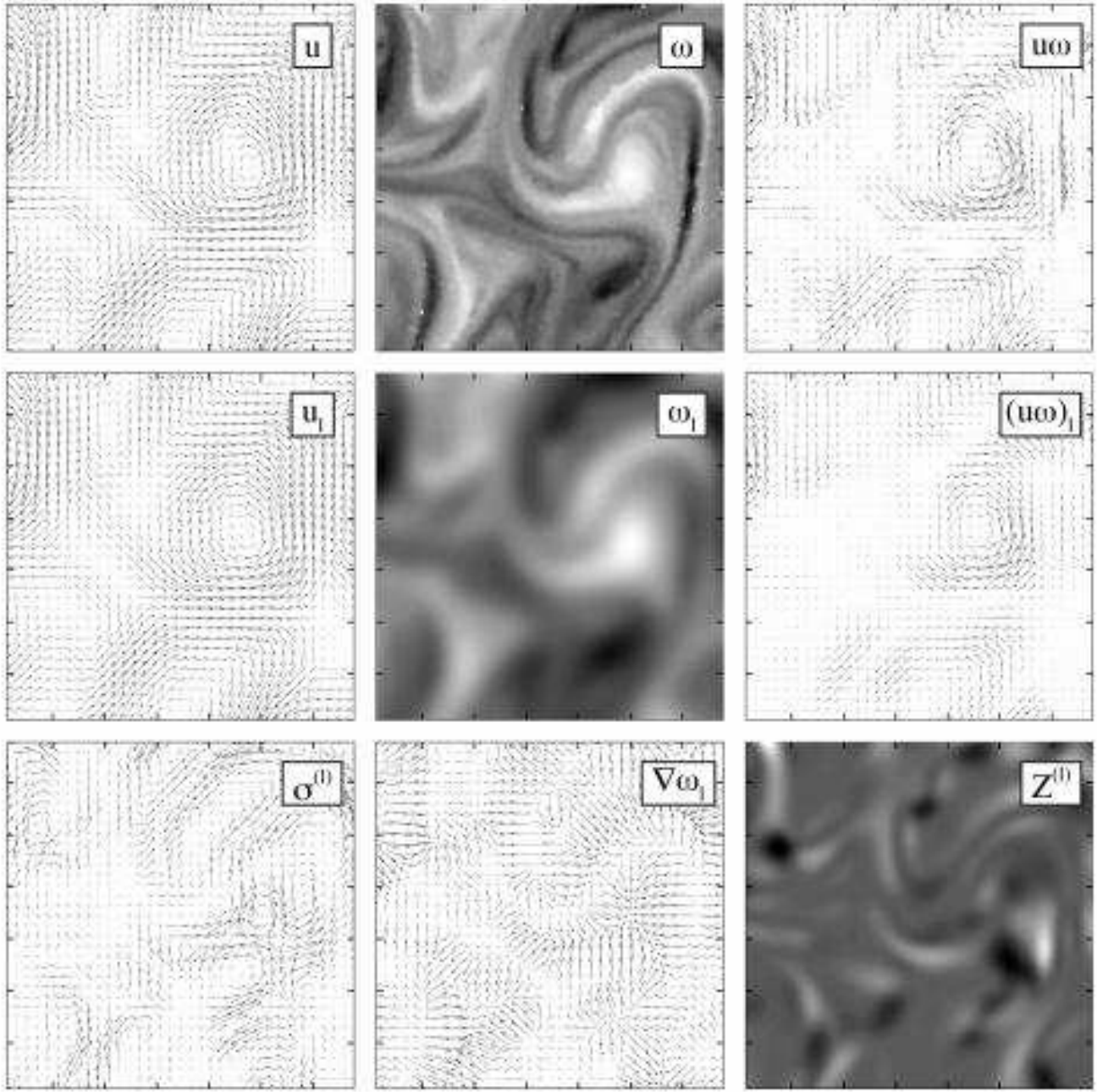}
   \caption{Obtaining the scale-to-scale enstrophy flux $Z^{(l)}$ for a 
     velocity field obtained from the soap film.  The filter function, $G_l$,
     used was Gaussian with $l=0.2$ cm. Row 1: Unfiltered velocity, ${\bf u}$,
     vorticity, $\omega$, and vorticity transport ${\bf u} \omega$. Row 2:
     Filtered velocity, ${\bf u}_l$, vorticity, $\omega_l$, and vorticity
     transport, $({\bf u} \omega)_l$.  Row 3: The subgrid vorticity transport
     vector ${\bf \sigma}^{(l)}$, large scale vorticity gradient $\nabla
     \omega_l$ and the scale-to-scale enstrophy transfer $Z^{(l)}$. }
\label{fig: enstrophy-flux-pictorial}
\end{figure*}

An almost identical method can be used to determine the scale-to-scale
enstrophy transfer of the flow.  The starting point, however, is the 2D Euler
equation for vorticity,
\begin{equation}
\frac{\partial \omega}{\partial t} + u_j \frac{\partial 
\omega}{\partial x_j}
= 0.
\label{eq: Euler-Vorticity}
\end{equation}
As above, the equation is contracted with a filter function, $G_l$, and the
coupling term is extracted,
\begin{eqnarray}
\frac{\partial \omega_l}{\partial t} + (u_j)_l\frac{\partial
   \omega_l}{\partial x_j} = -\frac{\partial
   \sigma^{(l)}_j}{\partial x_j},
\label{eq: Large-Scale-Euler-Vorticity}
\end{eqnarray}
where $\sigma^{(l)}_j = (u_j \omega)_l - (u_j)_l \omega_l$ is the subgrid
scale vorticity transport vector. Notice that the term $\sigma^{(l)}_j$
defined in Eq. \ref{eq: Large-Scale-Euler-Vorticity} has an almost identical
form to $\tau^{(l)}_{ij}$ in the energy equations. This general form is
typical of the filter approach. The coupling terms between large- and
small-scale fields often take the form $(ab)_l - a_l b_l$ for quadratic
nonlinearities.

To change the large-scale vorticity equation to an equation for large-scale
enstrophy evolution, one must multiply by $\omega_l$.  This yields
\begin{eqnarray}
\frac{\partial \Omega^{(l)}}{\partial t} &+& \frac{\partial}{\partial
   x_j}((u_j)_l\Omega^{(l)}) = -\omega_l\frac{\partial 
\sigma^{(l)}_j}{\partial
   x_j}\nonumber\\&=&-\frac{\partial}{\partial x_j}(\omega_l 
\sigma^{(l)}_j) +
   \sigma^{(l)}_j \frac{\partial \omega_l}{\partial x_j},
\label{eq: Large-Scale-Enstrophy-1}
\end{eqnarray}
where again the Leibniz rule was used to separate point-to-point transport
from inter-scale transfer.  Grouping the appropriate terms, as was done for
the energy equation, leads to the final form,
\begin{equation}
\frac{\partial \Omega^{(l)}}{\partial t} + \frac{\partial 
K^{(l)}_j}{\partial
   x_j} = -Z^{(l)},
\label{eq; Large-Scale-Enstrophy}
\end{equation}
where
\begin{eqnarray}
K^{(l)}_j &\equiv& (u_j)_l \Omega^{(l)} + \sigma^{(l)}_j \omega_l 
\label{eq: K-def}\\
Z^{(l)} &\equiv& -\sigma^{(l)}_j \frac{\partial \omega_l}{\partial x_j}
\label{eq: Zdef}
\end{eqnarray}
where the negative sign in front of the scale-to-scale coupling term,
$Z^{(l)}$, again makes down-scale transfer positive.

To make the above derivations more concrete, some examples are provided of
filtered fields.  For purposes of illustration, vorticity and enstrophy fields
are presented and an analysis of velocity and energy fields is left to a later
publication. Figure \ref{fig: enstrophy-flux-pictorial} displays steps in the
calculation of the scale-to-scale enstrophy transfer, $Z^{(l)}$, for a typical
vorticity field extracted from experimental data.  For this calculation the
filter function was a Gaussian with Fourier-space definition
\begin{equation}
G_{l}(\bf k)=e^{-\frac{|{\bf k}|^2}{k_l^2}},
\label{eq: GaussianKernel}
\end{equation}
where $k_l \equiv 2\pi/l$.  These figures illustrate that the application of
FA is straightforward: (1) compute the secondary field, $u_i\omega$, from the
measured velocity and vorticity fields; (2) perform a convolution of these
fields to obtain $\sigma_{i}^{(l)}$; (3) take the scalar product of
$\sigma_{i}^{(l)}$ with the appropriate gradient of the large-scale fields,
namely $\partial_i \omega_l$.

There is, however, a caveat for the general case. It may not always be
possible to separate the point-to-point transport from the scale-to-scale
transfer terms.  For the energy equation, the terms were determined by using
the Leibniz rule to separate out the Galilean invariant part of the energy
flux. For the enstrophy, the separation was obtained by analogy with the
energy equation rather than by a strict application of Galilean invariance.
There is no {\it a priori} expectation that such a separation will be as
simple, or even possible, for arbitrary nonlinear systems. In compressible
flows, for example, it is possible to measure the coupling terms, but the
point-to-point transport caused by small scales is tied to the scale-to-scale
transfer in a non-trivial way. The interpretation of such results must,
therefore, be done carefully.

\section{Systematic effects \label{sec:effects}}

This section will address the effect of varying the filter, $G_l$, on the
interpretation of the results, the consequences of the data being limited in
spatial extent ({\it i.e.} by boundaries), and variations in the results
caused by experimental or numerical limitations on the resolution of the data.

\subsection{Different Filters \label{subsec: Different Filters}}

\begin{figure}
   \includegraphics[width=3.2in]{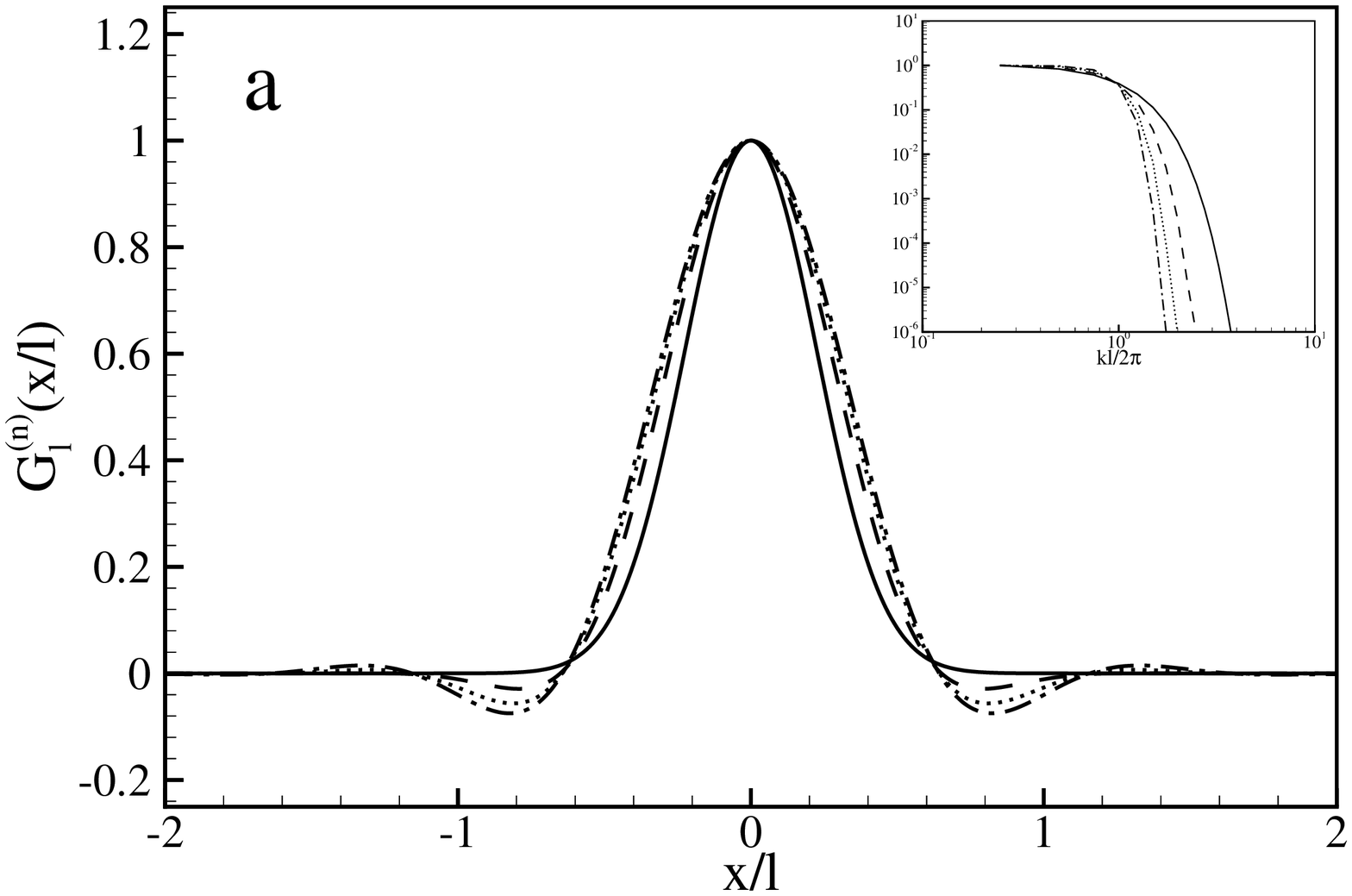}
   \includegraphics[width=3.2in]{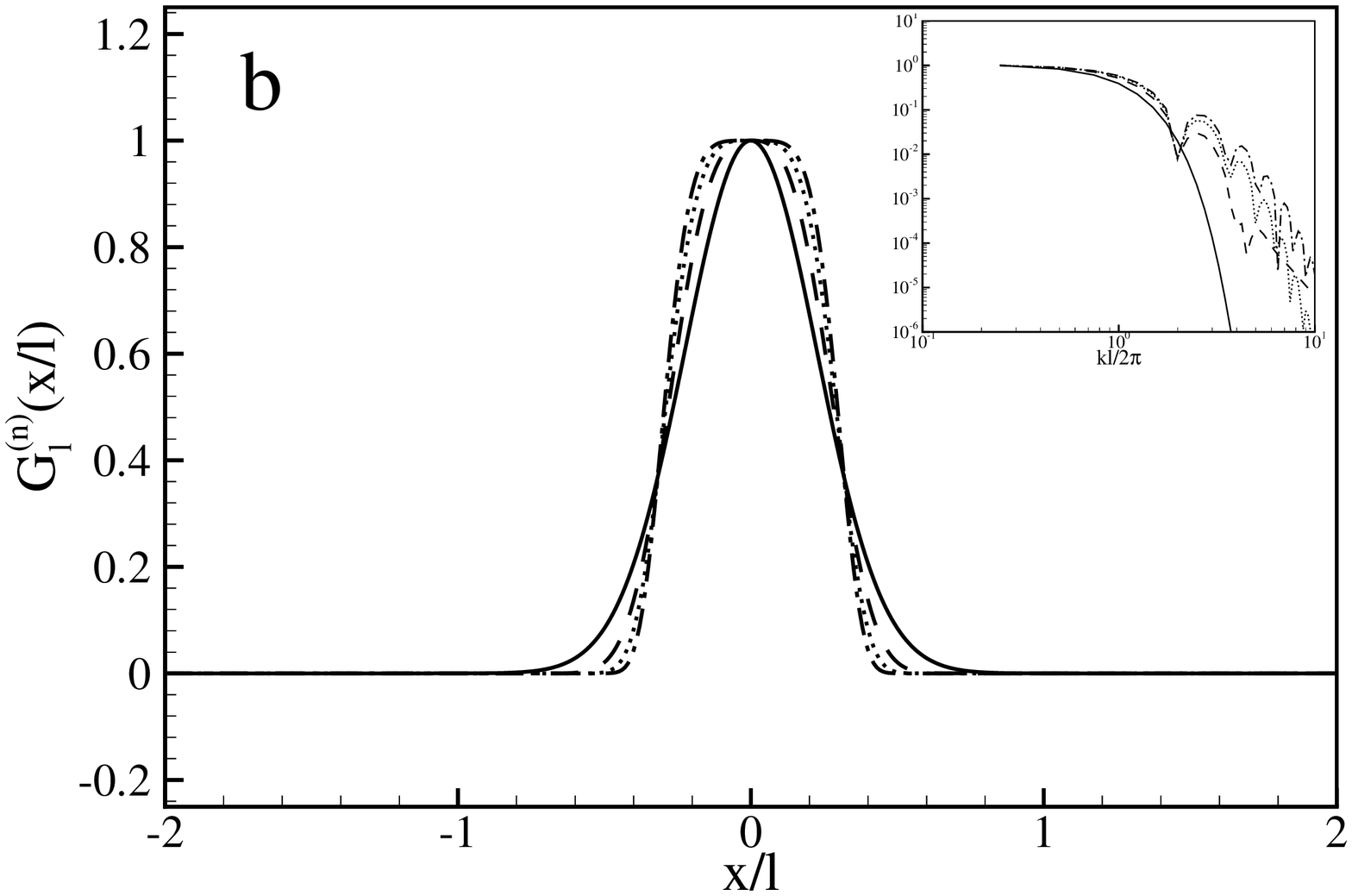}
   \caption{Illustration of the various filter functions $G^{(n)}_l$
     that are used in the paper (see Eq.~\ref{eq: sharpfourier} and
     Eq.~\ref{eq: sharpreal} for details) For both plots the real space
     convolution function is presented with the root of the Fourier spectrum
     inset. (a) Fourier filter convolution kernels, for orders n=2 (solid),
     n=3 (dash), n=4 (dotted), n=5 (dash-dot). (b) Real space filters for
     orders n=2 (solid), n=3 (dash), n=4 (dotted),n=5 (dash-dot).}
\label{fig: filter-kernels}
\end{figure}

\begin{figure*}
  \includegraphics[width=2.25in]{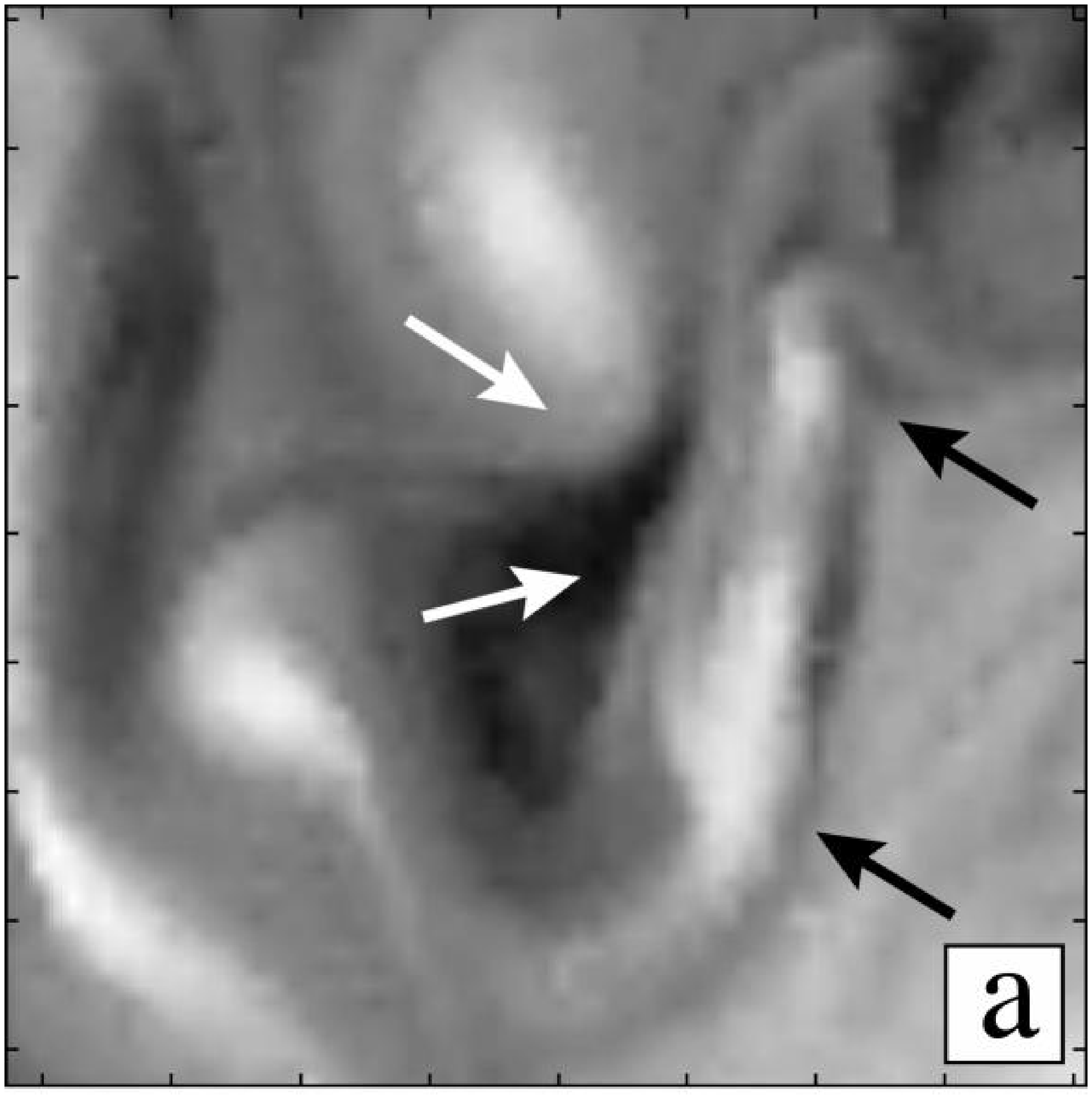}
  \includegraphics[width=2.25in]{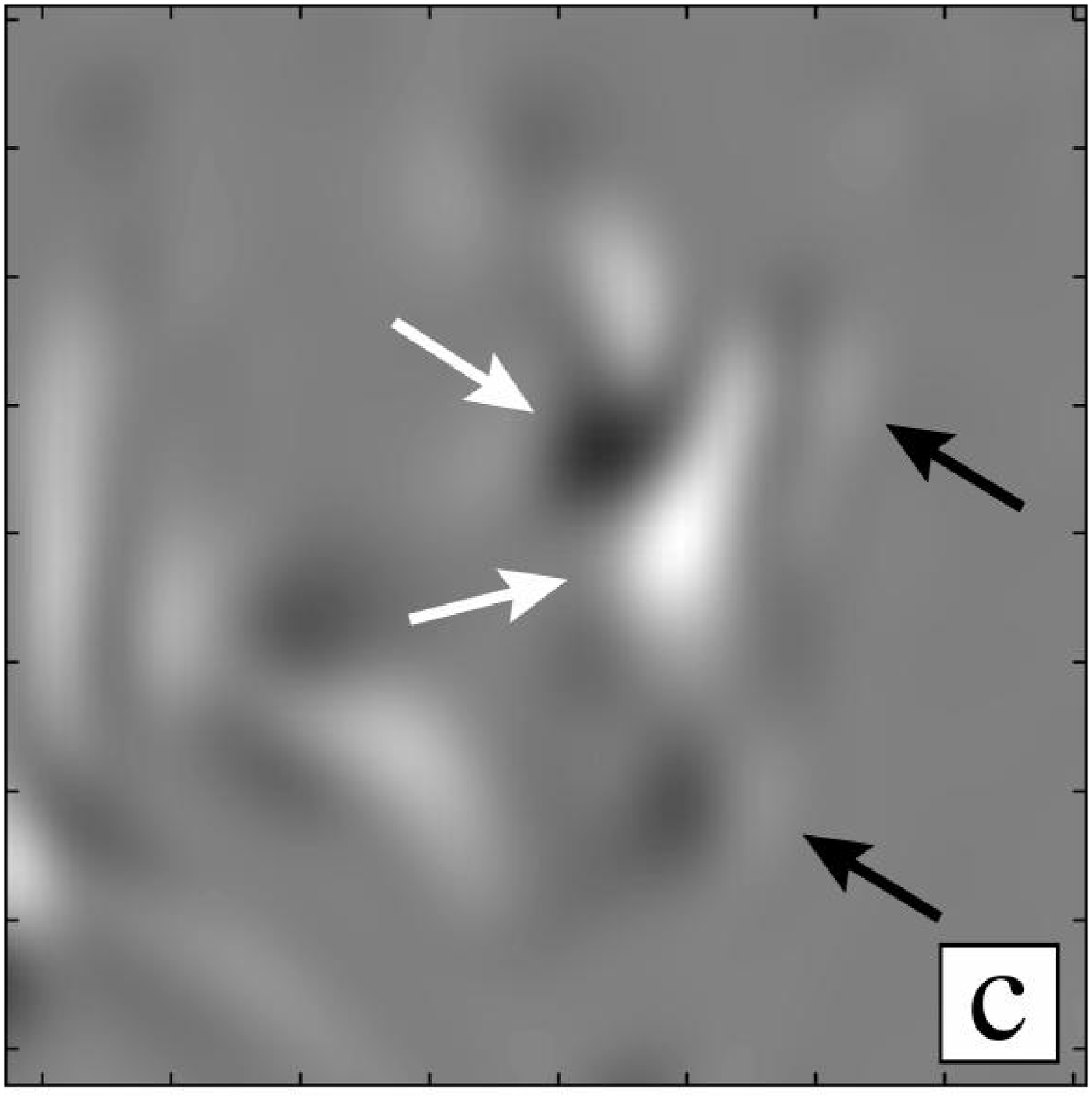}
  \includegraphics[width=2.25in]{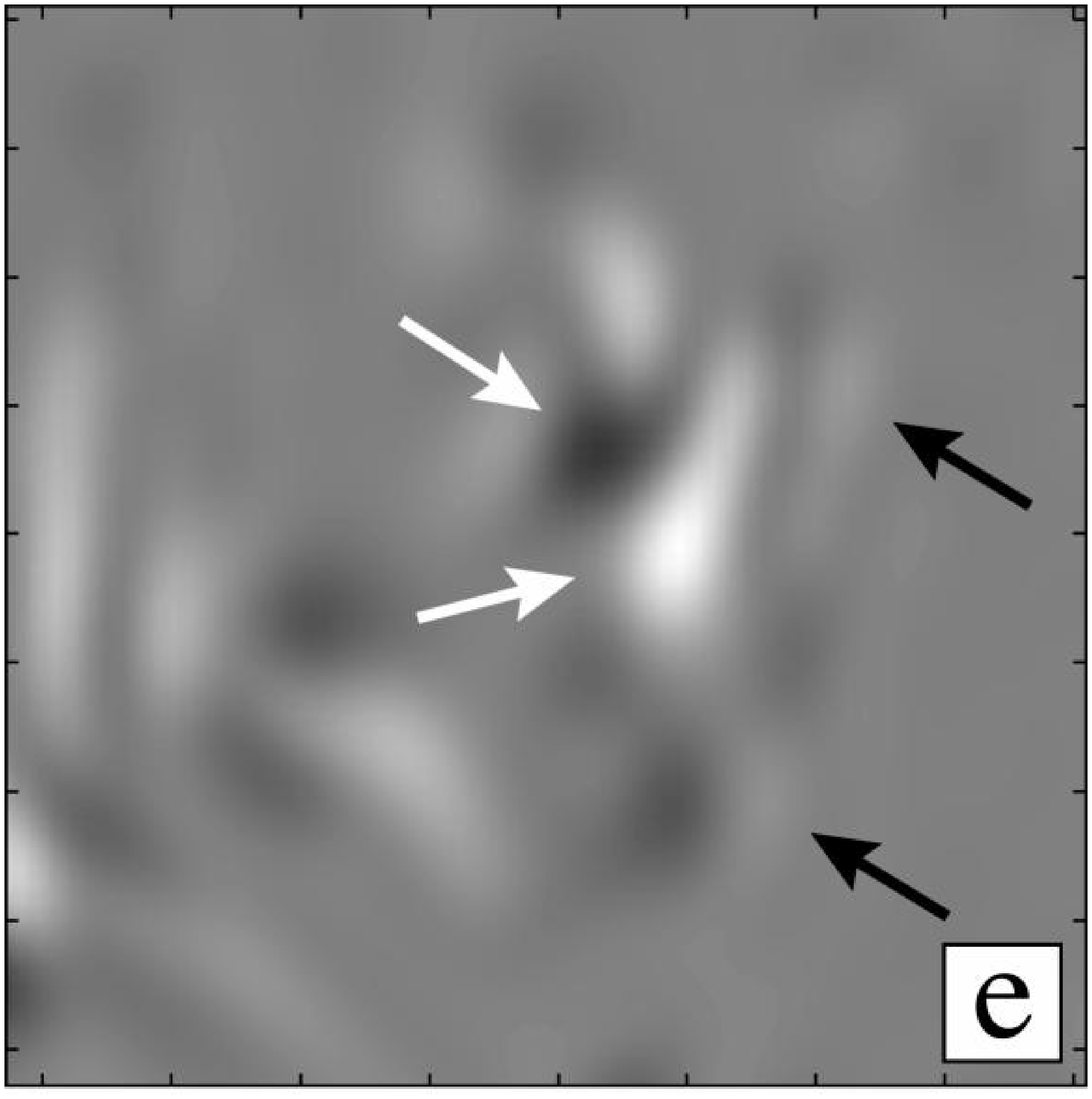}
  \includegraphics[width=2.25in]{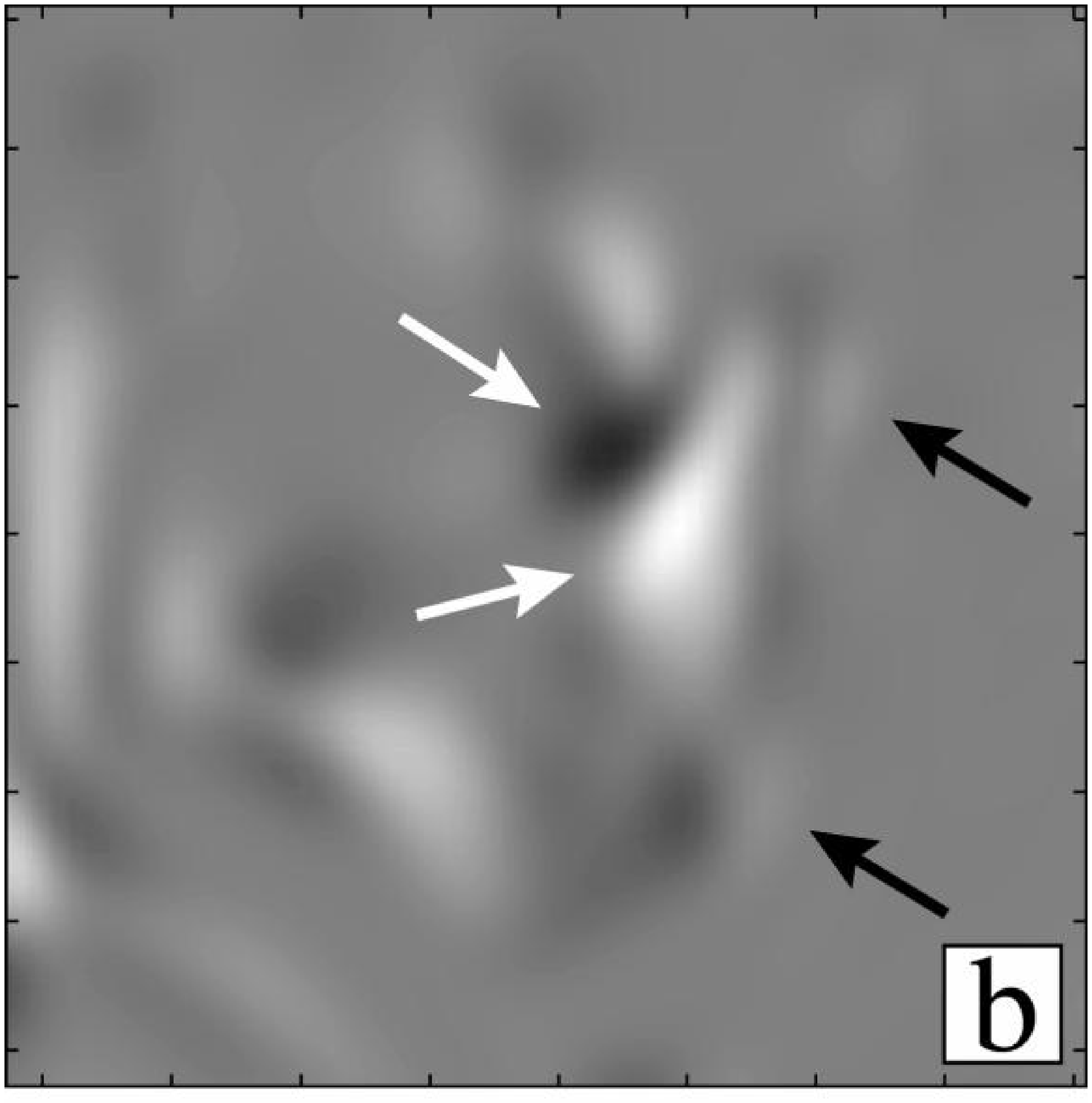}
  \includegraphics[width=2.25in]{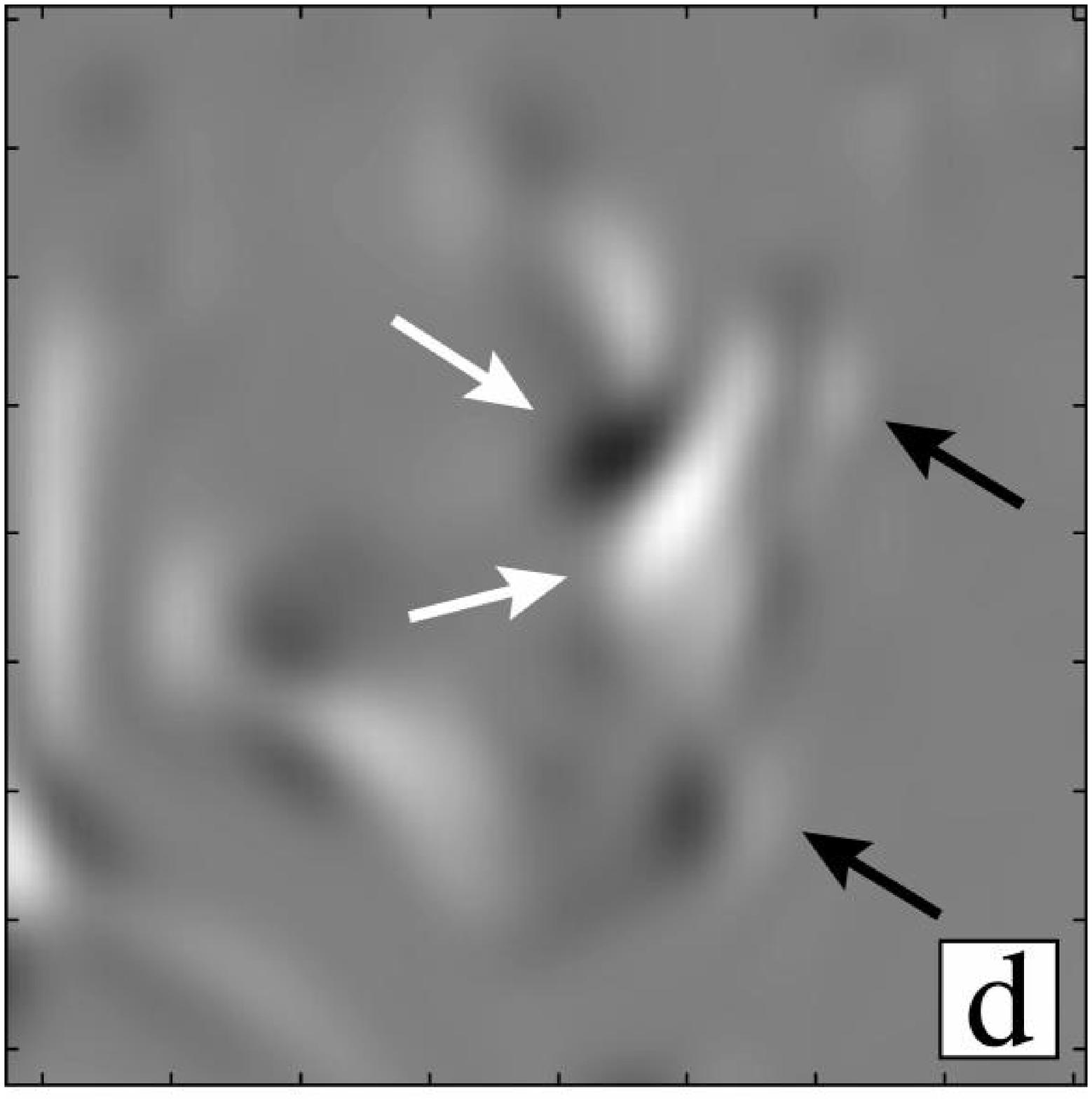}
  \includegraphics[width=2.25in]{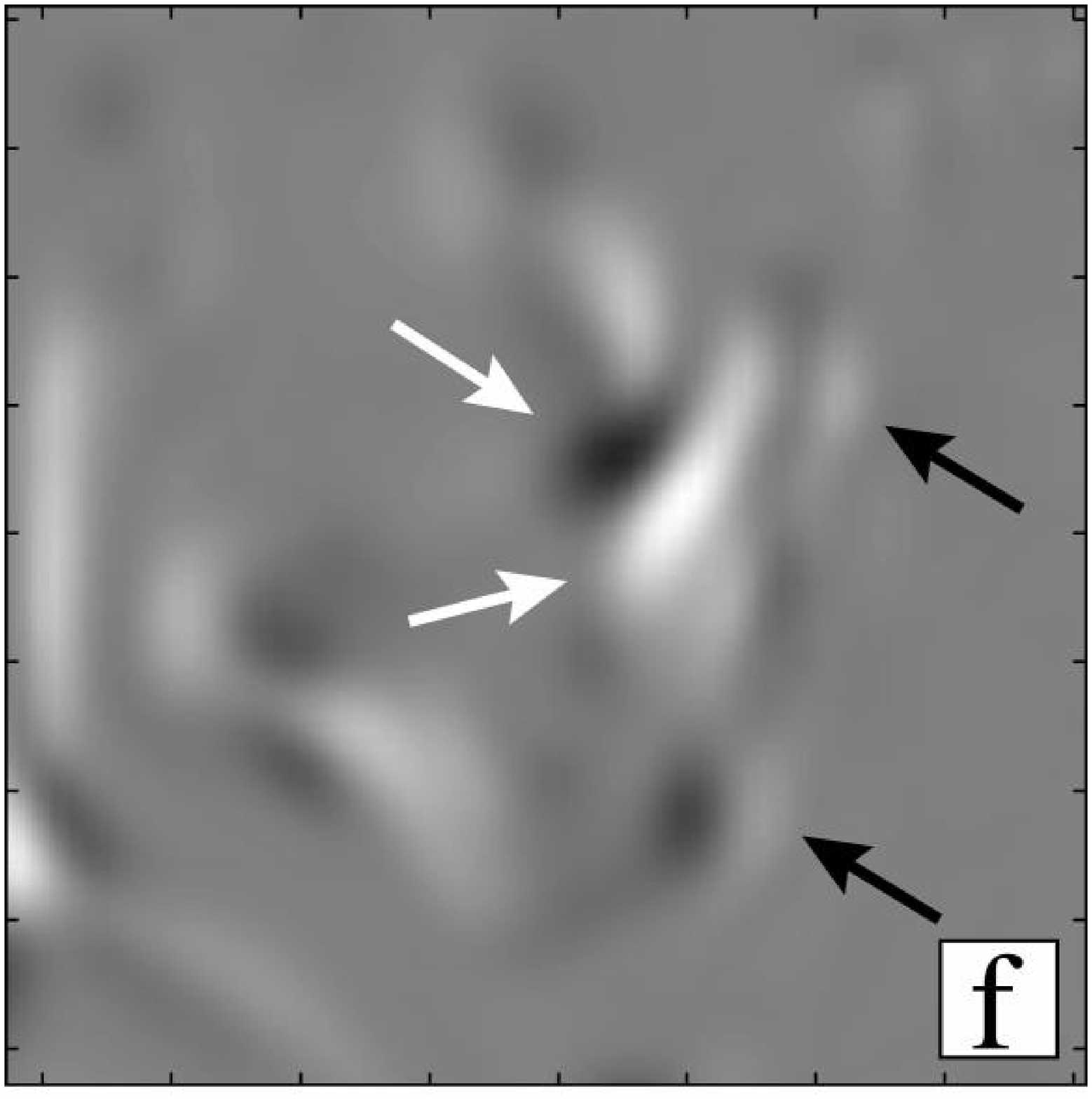}
  \caption{(a) Subsection of a typical vorticity field obtained from
    the soap film. (b)-(f) Calculation of $Z^{(l)}$ from vorticity field shown
    in (a) using $l=0.15$ cm and various Fourier and real space filters. (b)
    Gaussian filter (grayscale range $\pm 1.95 \times 10^7$s$^{-3}$) (c)
    Fourier filter of order $n=3$ (grayscale range $\pm 2.47 \times
    10^7$s$^{-3}$) (d) Real filter of order $n=3$ (grayscale range $\pm 1.93
    \times 10^7$s$^{-3}$) (e) Fourier filter of order $n=4$ (grayscale range
    $\pm 2.83 \times 10^7$s$^{-3}$) (f) Real filter of order $n=4$ (grayscale
    range $\pm 1.90 \times 10^7$s$^{-3}$).  The hatch marks represent $1$ mm
    increments.}
  \label{fig: enstrophyflux-ordered}
\end{figure*}

The interpretation of the results of FA may depend on the form of the filter
function, $G_l$. A Gaussian kernel was used in Eq.~(\ref{eq: GaussianKernel}),
but this particular choice was made only because it has a simple
interpretation in both real- and Fourier-space, taking the same form in both.
The FA technique imposes no such constraints in general. In some cases, one
may want to preferentially constrain the filter in real-space to a
well-defined length scale or, instead, may want to filter so as to select only
a sharp band of wave numbers. Either possibility can be explored using this
tool. It must be kept in mind, however, that sharpening the filter in real- or
Fourier-space causes a corresponding broadening of the filter in the other.
The impact and interpretation of varying the form of the filter on the
resulting fields is considered here.

The form of the Fourier filters used is
\begin{equation}
G^{(n)}_l(\bf k)=e^{-\left( \frac{|{\bf k}|}{k_l} \right)^{\it n}},
\label{eq: sharpfourier}
\end{equation}
where $n$ is the filter order.  The case $n=2$ is the Gaussian filter
considered earlier.  As $n$ increases, the filter sharpens around the Fourier
mode corresponding to filter length $l$.  Similarly, the real-space filters
are defined by:
\begin{equation}
G^{(n)}_l(\bf x)=Ae^{-\left( \frac{|{\bf x}|}{l/\pi} \right)^{\it n}}.
\label{eq: sharpreal}
\end{equation}
Note that for $n=2$, the real filter is equivalent to the Fourier filter.  As
the real filter is sharpened it approaches an area average over a box of
diameter $2l/\pi$.  Both of these sets of filters are shown in Fig. \ref{fig:
  filter-kernels}. Other types of filter are possible, but are not considered
here.

From an experimental point of view, the sharper real-space filters are more
attractive than the Fourier filters because the real-space envelope of the
Fourier filters grows as the order of the filter is increased.  Since
experimental data is invariably windowed to the cross-section of the
measurement apparatus, this means that sharper Fourier filters quickly grow to
interact with boundaries.  For sharp real filters this is not a problem: as
the filter increases in sharpness it becomes more spatially compact.  For the
purpose of comparing with theory, however, sharper Fourier filters approach
the ideal; see, for example, the discussion given in Frisch
\cite{Frisch_Turbulence} in which the fields are filtered by an infinitely
sharp cutoff in Fourier-space.

In Fig.~\ref{fig: enstrophyflux-ordered}, the results of using a Gaussian, two
low-order Fourier-space filters, and two low-order real-space filters in the
calculation of the enstrophy transfer are shown (the associated vorticity
field is also displayed).  Superficially, the fields are fairly similar. The
strength of the fluctuations, however, changes with filter (the grey scale
limits of the fields are noted in the captions).  The change in magnitude is
stronger for Fourier filters than real-space filters, with the $n=4$
Fourier-space filter experiencing $50\%$ larger fluctuations than for $n=2$.

The similarity in the fields seems to indicate that the {\em relative}
magnitudes of enstrophy flux remain more or less constant.  In particular, the
stronger values of flux associated with powerful vortices (middle top and
middle left) have almost identical forms, though there may be a slight
sharpening of the lobes for both higher order real-space and Fourier-space
filters (which, we note, have a quadrupolar form). Higher-order Fourier-space
filters increase the symmetry of many features, {\it i.e.} they appear less
ovular (see for example the lobes indicated by the white arrows in
Fig.~\ref{fig: enstrophyflux-ordered}), whereas the contrast between features
and the background is enhanced by higher-order real-space filters.  On the
whole, however, the qualitative features are fairly insensitive to the form of
filter used, maybe surprisingly so. One might expect that the beating of the
sharper Fourier kernel in real space would show up more strongly in the
fields.  For the stronger enstrophy transfer signals this does not appear to
be the case.

Weaker signals are more sensitive to the particular choice of filter.
Consider the weak lobe in the lower middle of the field indicated by the black
arrow in Fig.~\ref{fig: enstrophyflux-ordered} (or the one in the middle right
also indicated by a black arrow).  For the Gaussian filter, the lobe is barely
visible.  For sharper filters in both real- and Fourier-space ($n=4$),
however, the prominence of this lobe with respect to the stronger signals in
the system is enhanced.

The most important observation is that the qualitative structure of the field
is relatively unaffected by the choice of filter.  Sharper filters increase
the ``contrast'', but neither eliminate nor create structural features in the
flow. For the purposes of correlating scale-to-scale transfer to topological
features, this is an important feature of FA.

\begin{figure}
  \includegraphics[width=3.2in]{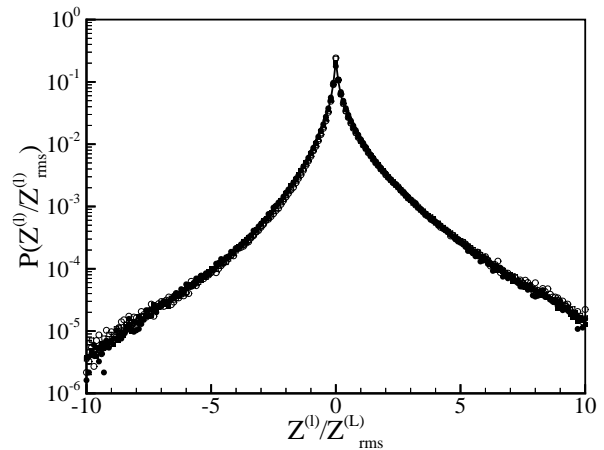}
  \caption{The probability distribution function (normalized by RMS) for
    $Z^{(l)}$ obtained using a Gaussian (solid line) two Fourier-space filters
    of order $n=3$(solid squares) and $n=4$ (solid circles) and two real-space
    filters of order $n=3$(open squares) and $n=4$ (open circles). The filter
    length was $l=0.2$ mm.}
  \label{fig: enstrophypdforder}
\end{figure}

The above comparison is only qualitative. A more accurate quantitative
comparison is presented in Fig.~\ref{fig: enstrophypdforder}, where the
probability distribution function of the enstrophy transfer is presented for
the same set of filters used in Fig.~\ref{fig: enstrophyflux-ordered}.  The
agreement between the different filters is quite good, but a little deceptive.
Note, first, that the magnitude of the RMS fluctuations has been normalized
out.  Second, there may be (though it is, perhaps, below the noise level) a
slight increase in asymmetry, in particular in the negative tail of the PDF
for real-space filters (the open symbols). Given that in Fig.~\ref{fig:
  enstrophyflux-ordered} the large values of enstrophy transfer, corresponding
to the tails of the PDF, are relatively insensitive to changes in the filter,
this collapse of the PDFs is reasonable.

\begin{figure}
  \includegraphics[width=3.2in]{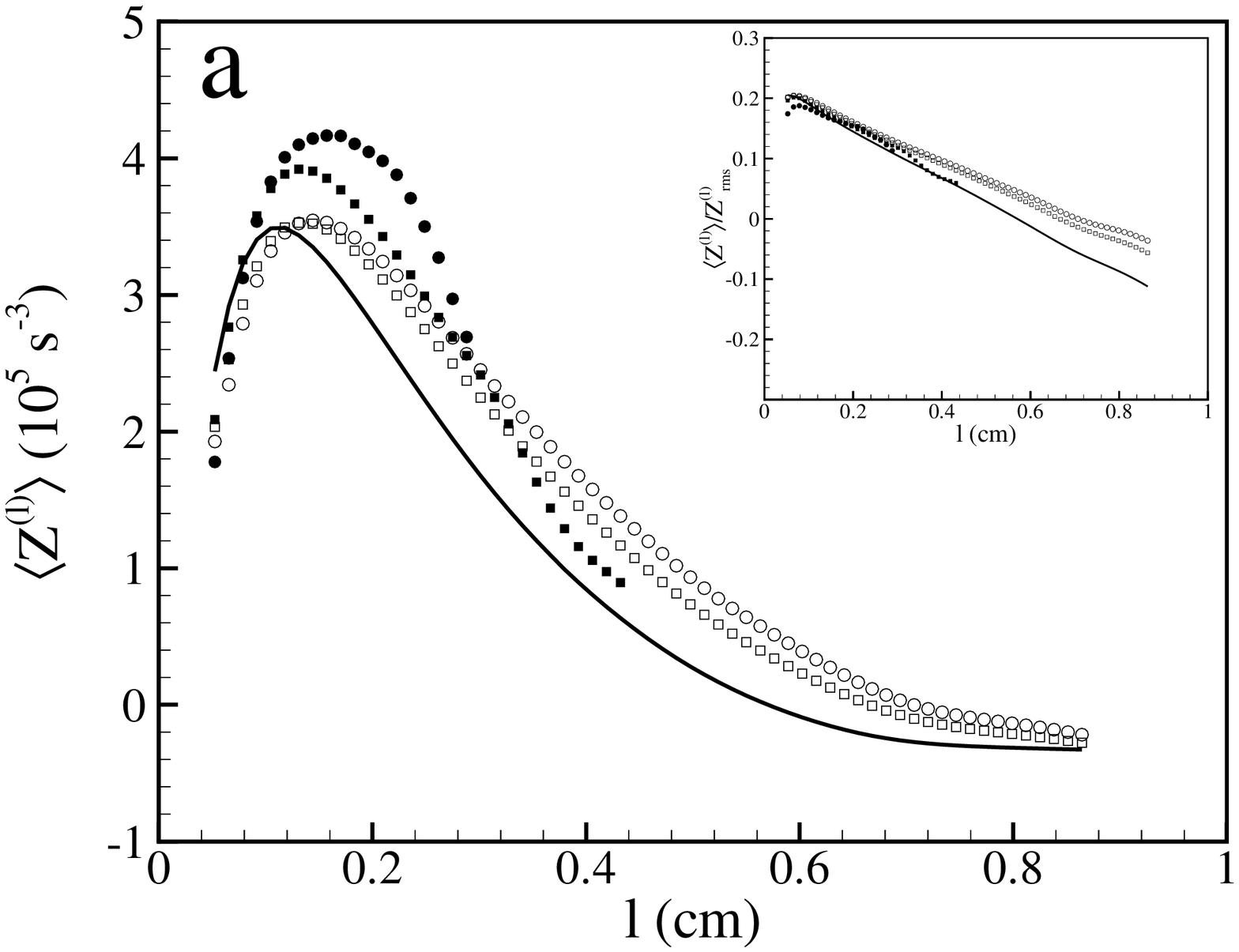}
  \includegraphics[width=3.2in]{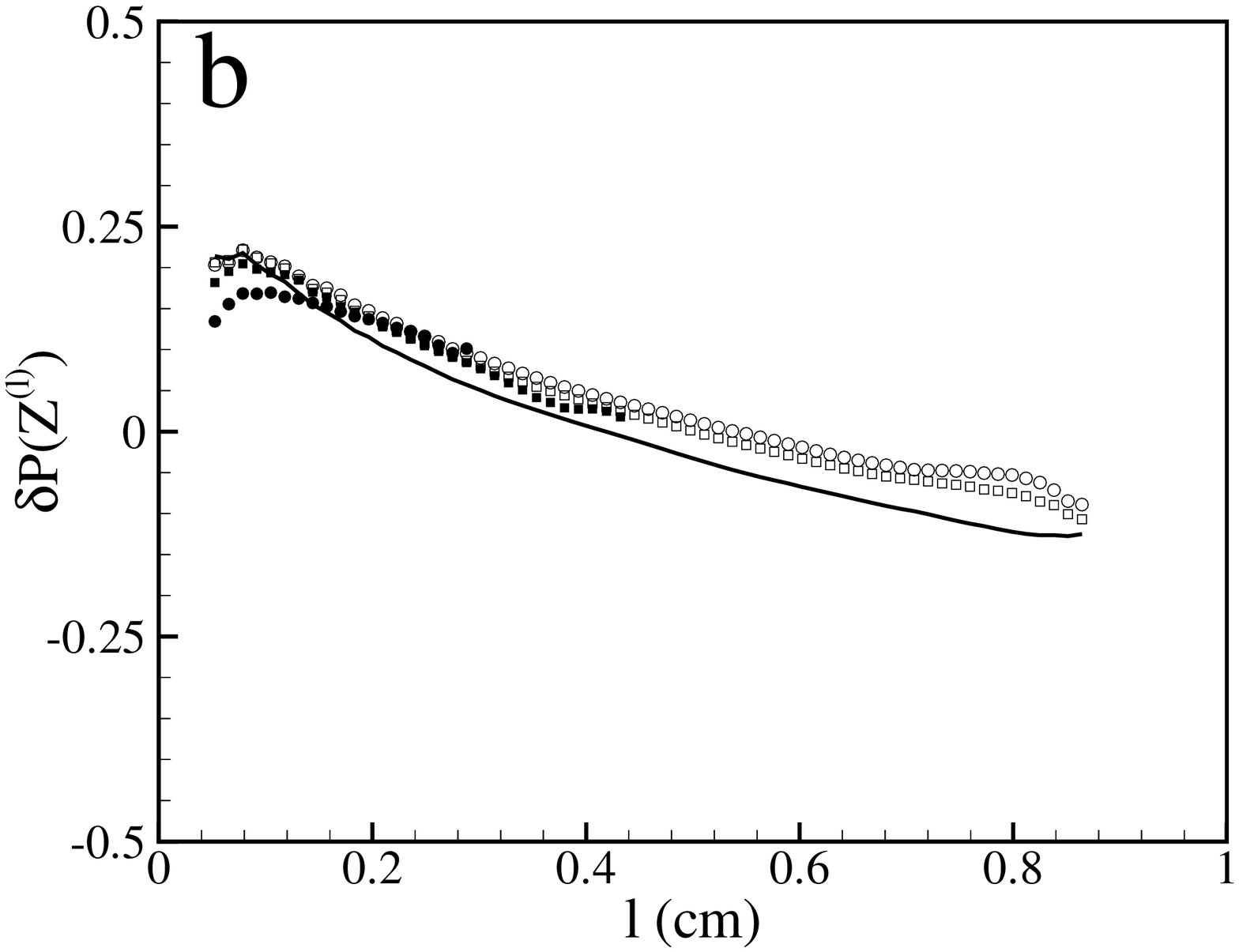}
  \caption{(a) The average enstrophy flux $\langle Z^{(l)} \rangle$
    and (b) $\delta P(Z^{(l)})$ (see text) for a range of length scales
    calculated using data from the soap film and a Gaussian filter function
    (solid line), two Fourier space filters of order $n=3$(solid squares) and
    $n=4$ (solid circles) and two real-space filters of order $n=3$ (open
    squares) and $n=4$ (open circles).  Inset in (a) is the average fluxes
    normalized by the RMS fluctuation size for the respective filters.}
  \label{fig: avgenstrophyorder}
\end{figure}

On the other hand, in Fig.~\ref{fig: enstrophyflux-ordered} the weaker values
of inter-scale transfer have somewhat increased contrast.  This change is
emphasized in the lowest-order moments of the distribution, rather than in the
tails. The average, $\langle Z^{(l)} \rangle$, and fractional sign-probability
comparison, $\delta P(Z^{(l)}) \equiv P(Z^{(l)}>0) - P(Z^{(l)} < 0)$, are
shown in Fig.~\ref{fig: avgenstrophyorder}. Here, there is a significant
difference in the average enstrophy flux as a function of filter order.  In
particular, the average rises and falls more sharply for the higher-order
Fourier-space filters than it does for the real-space filters. Also, the area
fraction saturates at a smaller value, then falls more quickly.  The signs of
the average flux and the area difference, however, are quite robust, although
the magnitudes seem to vary (in some places by a factor of two over this range
of filter orders). The rise in the peak of average scale-to-scale transfer is
reminiscent of ringing such as takes place in the Gibbs phenomenon.  It is not
clear whether or not this is the source of the change.

\subsubsection{Interpretation}

At this point, one might ask: for which filter is the result closest to the
``real'' enstrophy flux?  This question depends entirely on what one means by
``enstrophy flux''.  The scale-to-scale transfer between wave numbers is most
closely approximated by higher-order Fourier filters. Because of the
associated broadening of the filter in real-space, however, the resulting
fields are not as good a measure of the spatially-local enstrophy transfer.
On the other hand, for the flux to be localized in physical space for
comparison with real-space structures, sharp real-space filters are
preferable.  In this limit, the enstrophy flux can no longer be defined as the
movement of enstrophy from small Fourier modes to larger ones but is actually
a measure of the flux out of some bands and into others (which do not
necessarily have larger wave number).  The strength of FA does not lie in its
ability to measure {\em the exact magnitude of the transfer}, but rather that
the sign of the transport and the qualitative form of the fields is robust to
changes in the filter.

\subsection{Boundaries \label{subsec: Boundaries}}

An important consideration when using FA, one that is also a major issue in
LES, is the presence of boundaries (either physical or as limits of the
viewing window). As a Fourier-space filter grows in order, and correspondingly
in real-space extent, less and less of the data can be used as the boundaries
begin to affect the computation of the convolution in regions farther and
farther into the interior.

\begin{figure}
   \includegraphics[width=3.2in]{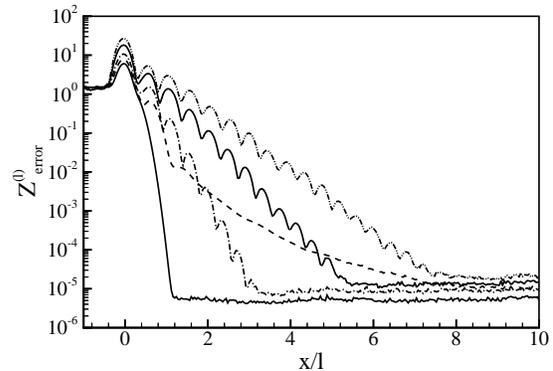}
   \caption{Interaction of various order Fourier-space filters with a
     boundary at $x/l=0$.  The graphs denote the RMS difference
     between the enstrophy flux measured in the presence of a boundary
     and that without.  The various curves are for a Gaussian filter
     (solid-line), and Fourier filters of order $n=3$ (dashed), $n=4$
     (dash-dot), $n=6$ (dotted), $n=8$ (dash-dot-dot).}
\label{fig: filter-convergence}
\end{figure}

To investigate the influence of finite system size a random periodic
streamfunction was generated on a $512\times512$ grid.  This streamfunction
was then used to obtain velocity and vorticity fields from which the enstrophy
flux field, $Z^{(l)}_0$, was computed.  The left half of the fields was then
set to zero and the enstrophy field, $Z^{(l)}_b$, recalculated.  The
normalized RMS difference,
\begin{equation}
Z^{(l)}_\text{error} = \frac{\langle (Z^{(l)}_b - Z^{(l)}_0)^2
\rangle^{1/2}}{ (Z^{(l)}_0)_\text{RMS}},
\end{equation}
is shown in Fig.~\ref{fig: filter-convergence} as a function of $x/l$, the
distance from the introduced boundary normalized by the filter size.  The
various plots are given for Fourier-space filters of different orders (see Eq.
\ref{eq: sharpfourier}).

With the exception of the order $3$ filter there is a continuous increase in
the boundary affects.  Taking a nominal error rate of $10^{-4}$ as acceptable,
a boundary of $b \approx l(n-1)$ is appropriately sized for the data. This
results in a loss of $2l(n-1)$ in linear box size since one must apply the
condition to left-right (top-bottom) boundaries. This effective boundary has
been adhered to throughout this paper. For real-space filters the boundaries
become less of an issue. Indeed, for the sharpest real-space filter (a simple
average over a circle), the boundary is $l/2$.

\subsection{Effects of Finite Resolution \label{subsec: Finite 
Resolution}}

\begin{figure}
  \includegraphics[width=3.2in]{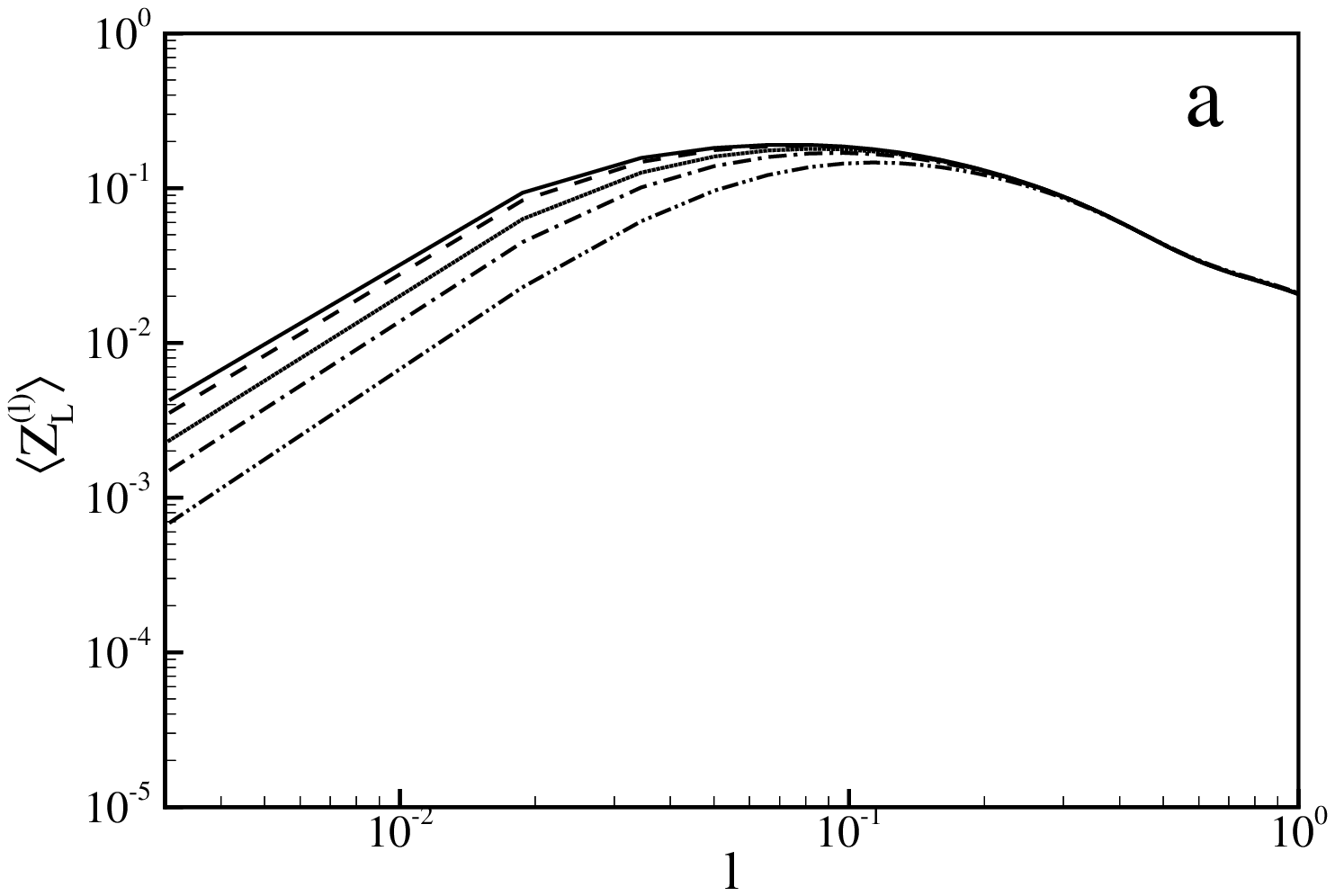}
  \includegraphics[width=3.2in]{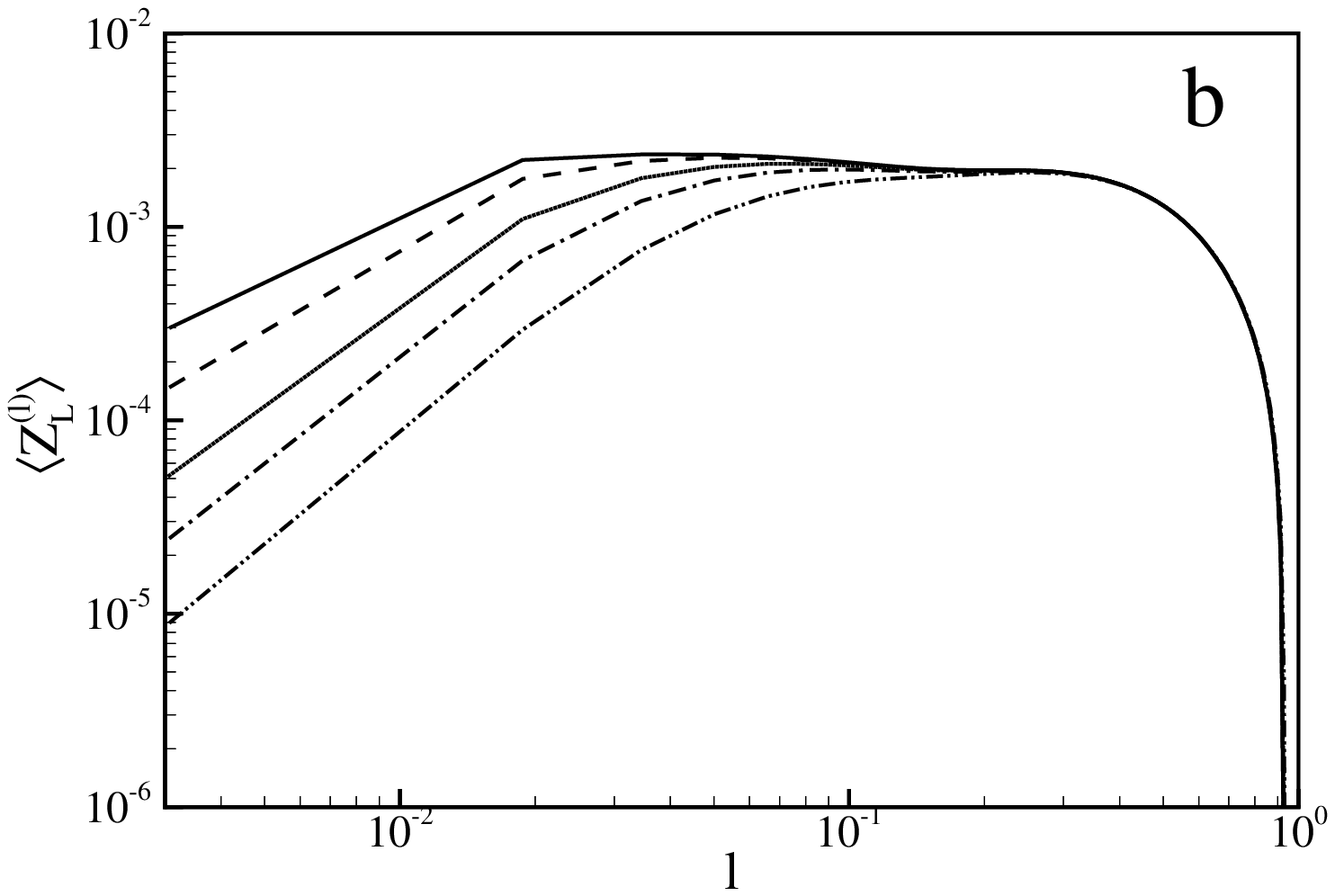}
  \caption{The average enstrophy flux calculated using numerical data
    for (a) Laplacian viscosity and (b) hyper-viscosity with a
    pre-filter of length $L$ = $0.01$ (dashed), $0.02$ (dotted),
    $0.03$ (dash-dot), $0.05$ (dash-dot-dot).  Also displayed is the
    total enstrophy flux (solid line) in the simulations.}
  \label{fig: resolutionanalysis}
\end{figure}

In this section the effect of finite measurement resolution on the ability
of FA to resolve the behavior of the enstrophy flux for a given filter
function is explored. In the preceding sections experimental data was used in
the analysis, but to estimate the effects of finite resolution it is necessary
to consider numerical data where the range of scales and the uncertainty in
the measured values is fairly well known. The numerical data also eliminates
any concerns about the effect of boundaries since the boundary conditions are
periodic.

Pre-filtering at the limit of the data resolution should result in no apparent
difference in the transfer from that computed from the original field. As the
pre-filter length, $L$, is increased above the resolution scale differences in
the measured transfer will increase. The question this raises is: how far
above $L$ are these finite-resolution effects felt?  To answer this question a
series of $20^{\rm th}$ order real-space filters with varying $L$ are applied
to two data sets simulating the local averaging inherent in experimental
measurements (such as image acquisition and particle tracking). From these
pre-filtered fields the enstrophy transfer, $Z_L^{(l)}$, is calculated and
compared with the full flux.  

The results of these calculations are shown for both data sets in
Fig.~\ref{fig: resolutionanalysis}. For both, the viscous cut-off scale is
between $0.01$ and $0.02$.  For pre-filters with $L$ at or below this length
there is little change in the magnitude of the average flux (slightly more
prominent in the case of hyper-viscosity). Once the filter size exceeds the
viscous scale, {\it i.e.}, penetrates at all into the inertial range, the
effects are quickly felt up to almost the injection scale, $0.4$. This may be
a general feature of FA, or it may be a peculiarity of the 2D enstrophy
transfer process being non-local.  In either case, it demonstrates the
importance of resolving the entire inertial range, including the viscous
scale, for an accurate representation of the enstrophy transfer to be
possible. Whether or not enstrophy transfer is local in Fourier-space is an
important question in its own right, and one which is sufficiently complex to
warrant exploration in a separate paper.

\section{Conclusions}

In the above discussion and use of FA, familiar turbulence assumptions, such
as homogeneity, isotropy, and inertial range, were not requirements of the
measurement. This is perhaps the greatest strength of FA: it can be applied to
any type of turbulent flow, regardless of that particular flow's properties.
All that is needed is the evolution equation for the system from which the
inter-scale transfer can be obtained.  The flow doesn't have to be turbulent,
FA can be applied to laminar or periodic flow and reasonable results obtained.
In other words, the interpretation of FA {\em does not rely on a pre-existing
  theory}.  Rather, it is a tool that can directly test notions of how
inter-scale transfer takes place in systems and can be used to build
appropriate theories.

This ``bottom up'' approach to physics (rather than a theoretical trickle
down) comes at a price: one must have highly resolved fields of data with a
significant range of spatial scales.  The former is a requirement imposed by
the resolution issues discussed above, which can only be relaxed if one
assumes a wavenumber local inter-scale transfer process.  The latter is a
limitation imposed by the data windowing and interaction of the filters with
the measurement boundary.  Of course, the standard techniques of measuring
velocity fields in fluids, namely particle imaging velocimetry or particle
tracking, more or less ensure that the measurements are not far from the
viscous scale, and thus of high enough resolution to not assume local
transfer.  This is because these techniques rely on groups of particles moving
coherently, which only holds when the scales being probed are small enough
that local Taylor expansions describe the flow.  The difficulty is in
simultaneously measuring a significant range of scales above the viscous scale
for meaningful information to be obtained.  With standard PIV practices this
is possible in 2D.  Only with holographic PIV is this possible in 3D. There,
however, the limitation is in the amount of data that can be obtained (order
10 fields is possible; currently, 1000 fields are not).

As discussed in the introduction, our ultimate intention for the FA technique
is to probe the mechanisms driving the inter-scale transfer in 2D turbulence,
with a particular eye towards the role of coherent structures in the transfer
process.  This will be done both via statistical analysis of fields in the
laboratory frame of reference (Eularian frame), as well as following the
motion of fluid parcels (Lagrangian frame).  FA, however, is not limited to
these measurements alone, but could find application to the inter-scale
transfer and mixing of passive scalars, or other quantities not necessarily
conserved in the inviscid limit ({\em i.e.} local topology).  And as hinted at
previously, FA can be applied (carefully!!!) to general non-linear systems
where an equation of motion is known and inter-scale transfer is of interest.

\section{acknowledgments}
We thank Greg Eyink, Phil Marcus, Misha Chertkov and Boris Shraiman for
interesting discussions and suggestions.

\end{document}